\documentclass[12pt]{article}

\usepackage{amsmath,amssymb,amsbsy,amsfonts,latexsym,graphicx}
\usepackage{color,array}
\usepackage[nospace,sort,compress]{cite}
\usepackage{emptypage} 
\usepackage[usenames,dvipsnames]{xcolor}
\usepackage[unicode,psdextra]{hyperref}
\hypersetup{
    colorlinks=true, 
    citecolor=ForestGreen,
    linkcolor=RoyalBlue,
    urlcolor=RoyalBlue,
    pdfstartview=FitV,
    breaklinks=true,
    linktocpage=true}

\usepackage{bookmark}

\numberwithin{equation}{section}

\allowdisplaybreaks

\evensidemargin \oddsidemargin

\begin{document}


\begin{titlepage}

\renewcommand{\thefootnote}{\fnsymbol{footnote}}


\begin{flushright}
\end{flushright}

\vspace{15mm}
\baselineskip 9mm
\begin{center}
  {\Large \bf Supersymmetric instantonic D1-branes \\
   in AdS$_5\times$S$^5$ background}
\end{center}

\baselineskip 6mm
\vspace{10mm}
\begin{center}
Jaemo Park\footnote{\tt jaemo@postech.ac.kr} and
Hyeonjoon Shin\footnote{\tt nonchiral@gmail.com}
\\[10mm]
  {\sl Department of Physics \&
       Center for Theoretical Physics,\\
       POSTECH, Pohang, Gyeongbuk 37673, South Korea}
\\ and \\
  {\sl Asia Pacific Center for Theoretical Physics, \\
       Pohang, Gyeongbuk 37673, South Korea}
\end{center}

\thispagestyle{empty}

\vfill
\begin{center}
{\bf Abstract}
\end{center}
\noindent
According to the covariant open superstring description of
1/2-BPS D-branes in the AdS$_5\times$S$^5$ background, there are two
kinds of purely instantonic D-branes.  One is the well known D(-1)-brane
or D-instanton, and another is the D1-brane spanning a two dimensional
subspace inside S$^5$.  We identify the actual 1/2-BPS instantonic
D1-brane configurations and their supersymmetry structures.
From the results, it is concluded that any spherical D1-instanton
with the radius equal to that of S$^5$ is 1/2-BPS.
After evaluating the 1/2-BPS D1-instanton action, we discuss the
relation of D1-instanton with the $(p, q)$ string instanton.  We
also discuss the possible 1/2-BPS and 1/4-BPS multiple D1-instanton
configurations.
\\ [15mm]

\vspace{5mm}
\end{titlepage}

\baselineskip 6.6mm
\renewcommand{\thefootnote}{\arabic{footnote}}
\setcounter{footnote}{0}

\tableofcontents

\section{Introduction}

The covariant open superstring description of D-branes
\cite{Lambert:1999id,Bain:2002tq} has been a useful way
in classifying supersymmetric D-branes, especially 1/2-BPS D-branes,
in a given supersymmetric background.
It has been successfully applied to some important backgrounds
in superstring theories such as the flat spacetime \cite{Lambert:1999id},
IIB plane wave \cite{Bain:2002tq}, IIA plane wave \cite{Hyun:2002xe},
AdS$_5 \times$ S$^5$ \cite{Sakaguchi:2003py,Sakaguchi:2004md,
ChangYoung:2012gi,Hanazawa:2016lvo}, and AdS$_4 \times \mathbf{CP}^3$
\cite{Park:2018gop} backgrounds.

The data obtained after the classification of supersymmetric D-branes
are however `primitive' in a sense that they do not tell us about
which configuration of a given D-brane is really supersymmetric and
which part of the background supersymmetry is preserved on the D-brane
worldvolume.  If the background spacetime geometry is for example
given by a product of some subspaces, the only information contained
in the data is the possible numbers of Neumann directions of open
superstring end point in each of the subspaces. In spite of this fact,
the data provide us an efficient guideline for further exploration of
supersymmetric D-branes, which allows us to avoid a brute force approach.
Indeed, in our previous work \cite{Park:2017ttx}, this has been
illustrated in the investigation of 1/2-BPS D-brane configurations in
the AdS$_5 \times$ S$^5$ background.

The work of \cite{Park:2017ttx} has been based on the result of
covariant open superstring description
\cite{Sakaguchi:2003py,Sakaguchi:2004md,ChangYoung:2012gi,Hanazawa:2016lvo}
given in Table \ref{tablebps} and focused on the Lorentzian D-branes.
By the way, as one can see from the table, there are two purely
Euclidean or instantonic D-branes.  One is the well known D(-1)-brane
or D-instanton \cite{Chu:1998in, Kogan:1998re, Bianchi:1998nk},
which has played an important role in the study of nonperturbative
aspects of IIB superstring theory in the AdS$_5 \times$ S$^5$ background.
Another is the D1-brane labeled by (0,2) which spans a two dimensional
subspace inside S$^5$.

As of now, compared to the D(-1)-brane, the 1/2-BPS instantonic D1-brane
looks somewhat exceptional and its role or nature is not obvious.
Since instantons are always important in understanding the
nonperturbative structure of a given theory,
it is certain that such object seems to deserve to be investigated for
further development of superstring theory in the AdS$_5 \times$ S$^5$
background.  In this paper, having the information (0,2) of Table
\ref{tablebps}, we will try to identify the 1/2-BPS instantonic
D1-brane configurations and their supersymmetry structures.
Partly, the present work may be regarded as the continuation of the
study carried out in \cite{Park:2020hgt} where 1/2-BPS instantonic
membrane configurations in the AdS$_4 \times$ S$^7 / \mathbf{Z}_k$
background have been successfully classified.

In the next section, we give a brief description of the
AdS$_5 \times$S$^5$ background and its Killing spinor.  The Euclidean
action for D1-brane with its basic structure is discussed in
Sec.~\ref{d1-action}.  The identification of 1/2-BPS instantonic
D1-brane configurations with their supersymmetric sturcutures is
worked out in Sec.~\ref{bpsd1}.  In Sec.~\ref{evalaction},
we evaluate the action values of the 1/2-BPS configurations and
try to show the relation between D1-instanton and $(p, q)$ string
instanton action based on \cite{Schwarz:1995dk,Witten:1995im}.
Finally, we discuss the possibility of
supersymmetric multiple D1-instantons in Sec.~\ref{concl}.

\begin{table}
\begin{center}
\begin{tabular}{c|cccccc}
\hline
  & D(-1) &D1 & D3 & D5 & D7 & D9 \\ \hline\hline
($n$,$n'$) & (0,0) &
\begin{tabular}{c} (2,0) \\ (0,2) \end{tabular} &
\begin{tabular}{c} (3,1) \\ (1,3) \end{tabular} &
\begin{tabular}{c} (4,2) \\ (2,4) \end{tabular} &
\begin{tabular}{c} (5,3) \\ (3,5) \end{tabular} &
-- \\
\hline
\end{tabular}
\caption{\label{tablebps} 1/2-BPS D-branes in the AdS$_5\times$S$^5$
background. $n$ ($n'$) represents the number of Neumann directions in
AdS$_5$ (S$^5$).}
\end{center}
\end{table}

\section{AdS$_5 \times$S$^5$ background and Killing spinor}
\label{bg}

In this section, we briefly describe the AdS$_5 \times$S$^5$ background
and its Killing spinor.  In the Poincar\'{e} patch coordinate system,
the metric of the AdS$_5 \times$S$^5$ geometry is given by
\begin{align}
\label{metric}
ds^2 = g_{\mu\nu} dX^\mu dX^\nu
     = \frac{R^2}{z^2}
    \left[ - (dx^{\hat{0}})^2 + (dx^{\hat{1}})^2 + (dx^{\hat{2}})^2
           + (dx^{\hat{3}})^2 + dz^2
    \right] + ds_{S^5}^2 \,,
\end{align}
where $ds_{S^5}^2$ is the metric of five sphere S$^5$ of radius $R$.
As usual, the common radius $R$ is written as
\begin{align}
R^4 = 4 \pi g_s N \ell_s^4 \,,
\label{adsrad}
\end{align}
where $g_s$ is the string coupling constant, $\ell_s$ the string
length scale, and $N$ the number of D3-branes leading to the above
geometry in the near-horizon limit.  As for the five sphere part,
one may parametrize it by using the usual four polar angles and one
azimuthal angle.  However, because instantonic D1-brane configurations
spanning two dimensional subspace inside S$^5$ are of our interest,
it is more convenient to take another parametrization in which
two dimensional substructures are manifest as follows:
\begin{align}
\label{s5metric}
ds_{S^5}^2 =
& R^2
 \left[
  d \alpha^2
  + \cos^2 \alpha (d \theta_1^2 + \sin^2 \theta_1 d \phi_1^2)
  + \sin^2 \alpha (d \theta_2^2 + \sin^2 \theta_2 d \phi_2^2)
 \right] \,,
\end{align}
where the ranges of angles are $0 \le \alpha \le \pi/2$,
$0 \le \theta_{1,2} \le \pi$, and $0 \le \phi_{1,2} \le 2\pi$.

From the metric (\ref{metric}) with (\ref{s5metric}), we choose the
zehnbein to be
\begin{align}
& e^{\hat{0}, \hat{1}, \hat{2}, \hat{3}} = \frac{1}{z} d x^{\hat{0},
      \hat{1}, \hat{2}, \hat{3}} \,, \quad
  e^z = \frac{dz}{z} \,,
\notag \\
& e^1 = R d \alpha \,,
\notag \\
& e^2 = R \cos \alpha d \theta_1 \,, \quad
  e^3 = R \cos \alpha \sin \theta_1 d \phi_1 \,,
\notag \\
& e^4 = R \sin \alpha d \theta_2 \,, \quad
  e^5 = R \sin \alpha \sin \theta_2 d \phi_2 \,.
\label{zehn}
\end{align}
Here, we have adopted hatted numbers and $z$ as the tangent space
index values for the AdS$_5$ part to distinguish them from those of
the S$^5$ part. With this convention, the ten dimensional tangent
space index (denoted by $A$, $B$, $\dots$) is denoted as
\begin{align}
A = ( m, z, a) \,, \quad m= \hat{0}, \hat{1}, \hat{2},\hat{3},
\quad a = 1,2,3,4,5 \,,
\end{align}
and the Ramond-Ramond (RR) five-form field strength, another constituent
of the AdS$^5 \times$S$^5$ background in addition to the metric
(\ref{metric}), is written as
\begin{align}
F_5 =
4 e^{\hat{0}} \wedge e^{\hat{1}} \wedge e^{\hat{2}} \wedge
  e^{\hat{3}} \wedge e^z
+
4 e^1 \wedge e^2 \wedge e^3 \wedge e^4 \wedge e^5 \,.
\label{rr5}
\end{align}

The AdS$^5 \times$S$^5$ background composed of (\ref{metric}) and
(\ref{rr5}) with (\ref{s5metric}) is maximally supersymmetric.
Its supersymmetry structure is encoded in the solution of the spacetime
Killing spinor equation.  The Killing spinor equation itself for the
AdS$_5 \times$S$^5$ background is given by
\begin{align}
\left(
    \nabla \delta^{IJ}
    + \frac{1}{2R} e^A \hat{\gamma} \Gamma_z \Gamma_A \tau_2^{IJ}
\right) \eta^J =0 \,,
\label{kse}
\end{align}
where$\nabla = d + \frac{1}{4} \omega^{AB} \Gamma_{AB}$,
\begin{align}
\tau_1 = \sigma_1 \,, \quad \tau_2 = i \sigma_2\,,
\end{align}
($\sigma_{1,2}$ are the usual Pauli matices.) and
\begin{align}
\hat{\gamma} \equiv \Gamma^{\hat{0}\hat{1}\hat{2}\hat{3}} \,.
\label{ghat}
\end{align}
The spacetime Killing spinors $\eta^I$ ($I=1,2$) as the solution of
(\ref{kse}) are two Majorana-Weyl spinors and taken to have
ten dimensional positive chirality in this work,
$\Gamma^{11} \eta^I = \eta^I$.

As demonstrated in
\cite{Claus:1998yw,Skenderis:2002vf}, the Killing spinor equation
(\ref{kse}) is solved rather easily if we split $\eta^I$ as
\begin{align}
\eta^I = \eta^I_+ + \eta^I_- \,,
\end{align}
where $\eta^I_\pm$ are defined by
\begin{align}
\eta^I_\pm = P^{IJ}_\pm \eta^J
\label{projeta}
\end{align}
with the projection operator
\begin{align}
P^{IJ}_\pm = \frac{1}{2} (\delta^{IJ} \pm \hat{\gamma} \tau_2^{IJ}) \,.
\label{proj}
\end{align}
In this splitting, we note that $\eta^1_\pm$ and $\eta^2_\pm$ are not
independent from each other because
\begin{align}
\eta^2_\pm = \mp \hat{\gamma} \eta^1_\pm \,.
\label{eta12}
\end{align}
Thus, to avoid this redundancy, it is convenient to define
\begin{align}
\eta_\pm \equiv \eta^1_\pm \,,
\label{eta}
\end{align}
to which $\eta^1$ and $\eta^2$ are related by
\begin{align}
\eta^1 = \eta_+ + \eta_- \,, \quad
\eta^2 = - \hat{\gamma} (\eta_+ - \eta_-) \,.
\end{align}
If we now use $\eta_\pm$, the solution of the Killing spinor equation
(\ref{kse}) is obtained as
\begin{align}
\eta_+ &= z^{-1/2} U
       (\epsilon_+ + \Gamma_m x^m \epsilon_-) \,, \notag \\
\eta_- &= z^{1/2} \Gamma_z U \epsilon_- \,,
\label{ks}
\end{align}
where $\epsilon_\pm$ are constant spinors and $U$ is a spinorial
function of five angles of S$^5$.

The function $U$ satisfies
\begin{align}
\left(
    d + \frac{1}{4} \omega^{ab} \Gamma_{ab}
    + \frac{1}{2R} e^a \Gamma_{za}
\right) U = 0 \,,
\label{ueq}
\end{align}
which can be read off from the Killing spinor equation (\ref{kse}).
If we took the standard parametrization of S$^5$, we could simply adopt
the solution of this equation obtained in \cite{Lu:1998nu}.  However,
since we take a different parametrization (\ref{s5metric}), we should
solve the equation. Now the necessary ingredients for solving
Eq.~(\ref{ueq}) are the S$^5$ part of zehnbein (\ref{zehn}) and
the corresponding spin connections computed as
\begin{gather}
\omega^{12} = \sin \alpha d \theta_1 \,, \quad
\omega^{13} = \sin \alpha \sin \theta_1 d \phi_1 \,,
\notag \\
\omega^{14} = - \cos \alpha d \theta_2 \,, \quad
\omega^{15} = - \cos \alpha \sin \theta_2 d \phi_2 \,,
\notag \\
\omega^{23} = - \cos \theta_1 d \phi_1 \,, \quad
\omega^{45} = - \cos \theta_2 d \phi_2 \,.
\end{gather}
Since the Eq.~(\ref{ueq}) is a first order differential equation,
it is not so difficult to solve it, and the resulting solution
$U$ is obtained as
\begin{align}
U = e^{ -\frac{1}{2} \alpha \Gamma_{z1} }
  e^{ - \frac{1}{2} \theta_1 \Gamma_{z2} }
  e^{ \frac{1}{2} \phi_1 \Gamma_{23} }
  e^{ \frac{1}{2} \theta_2 \Gamma_{14} }
  e^{ \frac{1}{2} \phi_2 \Gamma_{45} } \,.
\label{usol}
\end{align}

\section{Euclidean D1-brane action}
\label{d1-action}

Before moving on to the investigation of 1/2-BPS instantonic
D1-branes in the AdS$_5 \times$S$^5$ background, let us consider
the D1-brane action and its basic structure.
Since the instantonic D1-brane we are concerned about is
an object in the Euclidean spacetime, its action is the Euclidean
one given by\footnote{We note that the bosonic action is sufficient for
our purpose.}
\begin{align}
S = T \int d^2 \zeta \sqrt{ | G + \mathcal{F} | }
   - i g_s T \int ( C_{(2)} + \mathcal{F} C_{(0)} ) \,.
\label{action}
\end{align}
Here, $T$ is the D1-brane tension given in terms of the string coupling
constant $g_s$ and the string length scale $\ell_s$ ($\ell_s^2 = \alpha'$),
\begin{align}
T = \frac{1}{2 \pi g_s \ell_s^2} \,,
\label{d1tension}
\end{align}
and $| G + \mathcal{F} |$ is the determinant of the sum of two
objects which are the induced worldvolume metric\footnote{The
worldvolume indices $i,j$ take values of 1 and 2.}
\begin{align}
G_{ij} = \partial_i X^\mu \partial_j X^\nu g_{\mu\nu} \,,
\end{align}
and the combination of worldvolume gauge field strength $F_{ij}$ and the
induced NS-NS two-form gauge field $B^{NS}_{ij}$,
\begin{align}
\mathcal{F}_{ij} = 2 \pi \alpha' F_{ij} - B^{NS}_{ij} \,.
\end{align}
The background fields $C_{(0)}$ and $C_{(2)}$ are the induced
R-R zero and two-form gauge fields respectively.

As described in the last section, the AdS$_5 \times$S$^5$ background
does not have nontrivial profiles for $C_{(0)}$, $C_{(2)}$ and $B^{NS}$.
Thus we can eliminate these fields in the action (\ref{action}).
From the resulting action, the equation of motion for the worldvolume
gauge field is obtained as
\begin{align}
\partial_i
\left(
    \frac{\mathcal{F}_{ij}}{\sqrt{ |G+\mathcal{F}|}}
\right) = 0 \,.
\label{eom}
\end{align}
It is easy to solve this eqaution and we see that the solution can be
parametrized by a constant $\beta$ as
\begin{align}
\frac{B}{\sqrt{ |G+\mathcal{F}|}} = \sin \beta \,,
\label{emsol0}
\end{align}
where we have defined
\begin{align}
B \equiv \mathcal{F}_{12} = 2 \pi \alpha' F_{12} \,,
\label{bdef}
\end{align}
and $\beta$ has the range of $-\pi/2 \le \beta \le \pi/2$.
The solution (\ref{emsol0}) can be rewritten as
\begin{align}
B = \sqrt{|G|} \tan \beta \,,
\label{emsol}
\end{align}
and enables us to express the D1-brane action as
\begin{align}
S = \frac{T}{\cos \beta} \int d^2 \zeta \sqrt{|G|} \,.
\label{d1}
\end{align}

\section{1/2-BPS D1-instantons}
\label{bpsd1}

In this section, we identify the 1/2-BPS instantonic D1-brane
configurations based on the data from the covariant open string
description of D-branes in the AdS$_5 \times$S$^5$ background.
For the investigation of instantonic object, the spacetime is
taken to be Euclidean.  However, the Killing spinor $\eta_\pm$
of (\ref{ks}) with (\ref{usol}) have been obtained in the Lorentzian
signature.  Thus, for the expressions in the previous section,
we should take the Wick rotation
$x^{\hat{0}} \rightarrow -i x^{\hat{0}}$, under which\footnote{We
use the same definition for $\hat{\gamma}$ given in (\ref{ghat}).
Thus $\hat{\gamma}^2 = -1$ in the Lorentzian signature, while
$\hat{\gamma}^2 = +1$ in the Euclidean one.}
\begin{align}
\Gamma^{\hat{0}} \longrightarrow - i \Gamma^{\hat{0}} \,, \quad
\hat{\gamma} \longrightarrow - i \hat{\gamma} \,.
\label{wick}
\end{align}
In ten dimensional Euclidean spacetime, $\eta_\pm$ and
$\epsilon_\pm$ are no longer
Majorana-Weyl and become Weyl spinors. This change of nature, however,
does not make any annoying issue in the present work, because our
concern is the consistent projection operators acting on $\eta_\pm$
(strictly speaking $\epsilon_\pm$) which identify the 1/2-BPS
configurations.

Having the Killing spinor, the 1/2-BPS instantonic D1-brane
configurations can be investigated by using the usual equation
\cite{Bergshoeff:1997kr}
\begin{align}
\eta^I = \Gamma^{IJ} \eta^J \,,
\label{branesusy}
\end{align}
which is obtained by combining the spacetime supersymmetry and
the worldvolume $\kappa$-symmetry transformation.
The symbol $\Gamma$ represents the spinorial matrix appearing in the
$\kappa$ symmetry projection operator and satisfies
 $\Gamma^2 =1$ and $\mathrm{Tr} \, \Gamma = 0$.
Its explicit form for the D1-brane in the Euclidean spacetime is
\begin{align}
\Gamma^{IJ} =
\frac{i}{2 \sqrt{ |G + \mathcal{F}|}} \epsilon^{ij}
( \gamma_{ij} \tau_1^{IJ} + \mathcal{F}_{ij} \tau_2^{IJ} )\,,
\end{align}
where $\gamma_{ij} = \partial_i X^\mu \partial_j X^\nu e_\mu^A e_\nu^B
\Gamma_{AB}$.
Now, by acting the projection operator $P^{IJ}_\pm$ of (\ref{proj}) on
the above equation (\ref{branesusy}), we can express (\ref{branesusy})
in terms of
$\eta_\pm$ of (\ref{eta}) as\footnote{Due to Eq.~(\ref{wick}), the
projector $P^{IJ}_\pm$ of (\ref{proj}) changes to
$P^{IJ}_\pm = \frac{1}{2} (\delta^{IJ} \mp i \hat{\gamma} \tau_2^{IJ})$.
Because of this, the relation between $\eta^1_\pm$ and $\eta^2_\pm$ of
(\ref{eta12}) becomes $\eta^2_\pm = \pm i\hat{\gamma} \eta^1_\pm$.}
\begin{align}
\begin{pmatrix} \eta_+ \\ \eta_- \end{pmatrix} =
\frac{\hat{\gamma}}{2 \sqrt{ |G + \mathcal{F}|}} \epsilon^{ij}
\begin{pmatrix}
    - \mathcal{F}_{ij} & \gamma_{ij} \\
    - \gamma_{ij} & \mathcal{F}_{ij}
\end{pmatrix}
\begin{pmatrix} \eta_+ \\ \eta_- \end{pmatrix} \,.
\end{align}
We have two equations.  However, they are actually equivalent,
because the equation from the first row is obtained by acting
\begin{align}
1 + \frac{\hat{\gamma}}{2 \sqrt{ |G + \mathcal{F}|}} \epsilon^{ij} \mathcal{F}_{ij}
\end{align}
on the equation from the second row.
Thus it is enough to consider only one of two equations.
Here, we will take the equation from the first row,
\begin{align}
\eta_+ = \frac{\hat{\gamma}}{2 \sqrt{ |G + \mathcal{F}|}} \epsilon^{ij}
( \gamma_{ij} \eta_- -\mathcal{F}_{ij} \eta_+ ) \,.
\end{align}
Then, by plugging the Killing spinor (\ref{ks})
into this equation, we get
\begin{align}
\frac{1}{\sqrt{z}}( 1 + \hat{\gamma} \sin \beta)
( \epsilon_+ + \Gamma_m x^m \epsilon_- ) =
\frac{\sqrt{z} \cos \beta}{2 \sqrt{|G |} }  \hat{\gamma}
U^{-1} \epsilon^{ij} \gamma_{ij} \Gamma_z U \epsilon_- \,,
\label{susychk}
\end{align}
where $U$ is given in (\ref{usol}) and we have used (\ref{emsol0}) and
(\ref{emsol}).

The Eq.~(\ref{susychk}) is the key for identifying the 1/2-BPS instantonic
D1-branes.  Since the covariant open string description tells us
that the instantonic D1-branes embedded only in the S$^5$ of
AdS$_5 \times$ S$^5$ geometry may have the possibility of being
1/2-BPS, we will consider the D1-brane configurations each of which
spans a certain two dimensional subspace of the S$^5$ and is a point
in the AdS$_5$ space.  Thus the coordinates $(z, x^m)$ in (\ref{susychk}),
the position of D1-brane in the AdS$_5$ space, are taken to be
constants and the $U$ has a specific expression corresponding to a given
configuration.  If we evaluate the right hand side of (\ref{susychk})
for a given D1-brane configuration and obtain the result which does not
depend on any worldvolume coordinate, then the configuration is
confirmed to be 1/2-BPS.\footnote{One may argue that any D1-brane
embedded in the S$^5$ is always 1/2-BPS if it is placed at $z=0$ or
$\infty$.  However, it seems that the object at such position should
be handled with care because such supersymmetry
structure may not be obtained by taking the limit $z \rightarrow 0$
or $\infty$ after solving (\ref{susychk}).
In our study, it is presumed that the position of
D1-brane in the AdS$_5$ space is generic and Eq.~(\ref{susychk})
is solved preferentially before taking any limit.}

Specifying a D1-brane configuration or embedding in S$^5$ is to choose
a static gauge for the worldvolume reparametrization.  From the five
sphere metric (\ref{s5metric}), we can figure out largely two types of
static gauge fixing conditions according to whether or not the coordinate
$\alpha$ is transverse to the D1-brane worldvolume.  We first investigate
the cases where $\alpha$ is a transverse direction.

Perhaps, the most immediate one would be the D1-brane that wraps a two
sphere parametrized by ($\theta_1, \phi_1$) or ($\theta_2, \phi_2$).
If we choose the static gauge as
\begin{align}
\zeta^1 = \theta_1 \,, \quad \zeta^2 = \phi_1 \,,
\label{config1}
\end{align}
with constant $\alpha$ and $\theta_2 = \phi_2 = 0$, then
$\sqrt{|G|} = R^2 \cos^2 \alpha \sin \theta_1$ and $U$ of (\ref{usol})
reduces to
\begin{align}
U = e^{ -\frac{1}{2} \alpha \Gamma_{z1} }
  e^{ - \frac{1}{2} \theta_1 \Gamma_{z2} }
  e^{ \frac{1}{2} \phi_1 \Gamma_{23} } \,.
\end{align}
The right hand side of (\ref{susychk}) is evaluated by using the following
computation:
\begin{align}
\frac{1}{2 \sqrt{|G |} } U^{-1} \epsilon^{ij} \gamma_{ij} \Gamma_z U
&= U^{-1} \Gamma_{23} \Gamma_z U
\notag \\
&= \left(
     \cos \alpha
      + \sin \alpha e^{ - \frac{1}{2} \phi_1 \Gamma_{23} }
           e^{ \theta_1 \Gamma_{z2} }
           e^{ \frac{1}{2} \phi_1 \Gamma_{23} }
   \right)
   \Gamma_{23} \Gamma_z \,.
\end{align}
We see that this expression depends explicitly on the
worldvolume coordinates $\theta_1$ and $\phi_1$.  Though this seems
to make the present configuration non-supersymmetric,
we can get rid of the dependence by taking the transverse
position of D1-brane in the $\alpha$ direction to be $\alpha = 0$,
at which the size of the S$^2$ parametrized by ($\theta_1, \phi_1$)
is maximized as one can see from (\ref{s5metric}).  Then, the
Eq.~(\ref{susychk}) becomes
\begin{align}
( 1 + \hat{\gamma} \sin \beta) ( \epsilon_+ + \Gamma_m x^m \epsilon_- ) =
z \cos \beta  \hat{\gamma} \Gamma_{23} \Gamma_z \epsilon_- \,.
\label{d1s2}
\end{align}
If we now introduce a projection operator
\begin{align}
P^x_{\pm} = \frac{1}{2} ( 1 \pm \hat{\gamma} ) \,,
\label{xproj}
\end{align}
and act on (\ref{d1s2}), we finally get
\begin{align}
\epsilon_{++}
&= \frac{\cos \beta}{1 + \sin \beta} z \Gamma_{23} \Gamma_z \epsilon_{-+}
  - \Gamma_m x^m \epsilon_{--} \,,
\notag \\
\epsilon_{+-}
&= - \frac{\cos \beta}{1 - \sin \beta} z \Gamma_{23} \Gamma_z \epsilon_{--}
  - \Gamma_m x^m \epsilon_{-+} \,,
\label{d1s2res}
\end{align}
where we have defined
\begin{align}
\epsilon_{+ \pm} \equiv P^x_\pm \epsilon_+ \,, \quad
\epsilon_{- \pm} \equiv P^x_\pm \epsilon_-  \,.
\label{epm}
\end{align}
The Eq.~(\ref{d1s2res}) clearly shows that $\epsilon_{+ \pm}$ is
given in terms of $\epsilon_{- \pm}$, while $\epsilon_{- \pm}$ are still
free parameters.  Thus it is concluded that the D1-brane wrapping
the S$^2$ parametrized by ($\theta_1$, $\phi_1$) placed at
$\alpha = 0$ is 1/2-BPS.  We note that this result continues to hold
for the D1-brane wrapping another S$^2$ parametrized by
($\theta_2$, $\phi_2$).  The differences are that $\Gamma_{23}$ in
(\ref{d1s2res}) is to be replaced by $\Gamma_{45}$ and the position
in the $\alpha$ direction should be $\alpha = \pi/2$.

The second case is the configuration corresponding to
the following static gauge,
\begin{align}
\zeta^1 = \phi_1 \,, \quad \zeta^2 = \phi_2 \,,
\label{t2}
\end{align}
with constant $\alpha$, $\theta_1$ and $\theta_2$.  In this gauge
choice, $\sqrt{|G|} = R^2 \cos^2 \alpha \sin \alpha \sin \theta_1
\sin \theta_2$ and $U$ is just given by (\ref{usol}).  Then, the
important part of the right hand side of (\ref{susychk}) is
computed as
\begin{align}
\frac{1}{2 \sqrt{|G |} } U^{-1} \epsilon^{ij} \gamma_{ij} \Gamma_z U
&= U^{-1} \Gamma_{35} \Gamma_z U
\notag \\
&=
\bigg(
   \cos \alpha e^{ - \phi_2 \Gamma_{45} }
   e^{ -\frac{1}{2} \phi_1 \Gamma_{23} }
   e^{ \theta_1 \Gamma_{z2} }
   e^{ -\frac{1}{2} \phi_1 \Gamma_{23} }
\notag \\
& + \sin \alpha \Gamma_{z1}
   e^{ - \frac{1}{2} \phi_2 \Gamma_{45} }
   e^{ \theta_2 \Gamma_{14} }
   e^{ - \frac{1}{2} \phi_2 \Gamma_{45} }
   e^{ - \phi_1 \Gamma_{23} }
\bigg) \Gamma_{35} \Gamma_z \,.
\end{align}
We see that there are explicit dependencies on the worldvolume coordinates
$\phi_1$ and $\phi_2$.  Unlike the previous case, one cannot
eliminate such dependencies for any choice of  $\alpha$, $\theta_1$
and $\theta_2$.  Therefore, the only solution of (\ref{susychk}) is
$\epsilon_\pm = 0$ and thus the D1-brane wrapping $\phi_1$ and $\phi_2$
is not supersymmetric.

The third case, the last one in which $\alpha$ is a transverse direction,
is the D1-brane wrapping the `diagonal' S$^2$ composed of two
S$^2$'s parametrized by $\theta_{1,2}$ and $\phi_{1,2}$.\footnote{This
configuration is inspired by that of instantonic D2-brane
\cite{Drukker:2011zy} in the AdS$_4 \times \mathbf{CP}^3$ background.}
If we let
\begin{align}
\vartheta = \theta_1 = \theta_2 \,, \quad
\varphi_\pm = \phi_1 = \pm \phi_2 \,,
\label{diagvar}
\end{align}
the corresponding static gauge is
\begin{align}
\zeta^1 = \vartheta \,, \quad
\zeta^2 = \varphi_+ \; ( \mathrm{or} \; \varphi_-)  \,,
\label{gaugediag}
\end{align}
with constant $\alpha$.  Let us first consider the configuration
where $\zeta^2 = \varphi_+$.   It leads us to have
$\sqrt{ |G| } = R^2 \sin \vartheta$ and
\begin{align}
U = e^{ -\frac{1}{2} \alpha \Gamma_{z1} }
  e^{ - \frac{1}{2} \vartheta ( \Gamma_{z2} - \Gamma_{14} ) }
  e^{ \frac{1}{2} \varphi_+ ( \Gamma_{23} + \Gamma_{45} ) } \,,
\end{align}
from (\ref{usol}), which in turn are used to get
\begin{align}
\frac{1}{2 \sqrt{|G |} } U^{-1} \epsilon^{ij}
\gamma_{ij} \Gamma_z U
&= U^{-1} \Gamma_{23} e^{\alpha (\Gamma_{24} + \Gamma_{35})}
   \Gamma_z U
\notag \\
&= \Gamma_z \Gamma_{23}
  e^{\alpha (\Gamma_{24} + \Gamma_{35} - \Gamma_{z1} ) } \,.
\label{diap}
\end{align}
Obviously, the last expression in (\ref{diap}) is independent
from any of the worldvolume coordinates and hence implies that
the present configuration is 1/2-BPS.  Then, after a bit of manipulation
with (\ref{diap}), Eq.~(\ref{susychk}) determining the
supersymmetry structure becomes
\begin{align}
\epsilon_{++}
&= \frac{\cos \beta}{1 + \sin \beta} z \Gamma_z \Gamma_{23}
  e^{\alpha (\Gamma_{24} + \Gamma_{35} - \Gamma_{z1} ) } \epsilon_{-+}
  - \Gamma_m x^m \epsilon_{--} \,,
\notag \\
\epsilon_{+-}
&= - \frac{\cos \beta}{1 - \sin \beta} z\Gamma_z \Gamma_{23}
  e^{\alpha (\Gamma_{24} + \Gamma_{35} - \Gamma_{z1} ) } \epsilon_{--}
  - \Gamma_m x^m \epsilon_{-+} \,,
\label{d1diap}
\end{align}
where $\epsilon_{+\pm}$ and $\epsilon_{-\pm}$ are defined in
(\ref{epm}) in terms of the projection operator (\ref{xproj}).
These equations clearly show that the
D1-brane wrapping the `diagonal' S$^2$ is 1/2-BPS.  Compared to
the previous 1/2-BPS configuration (\ref{config1}), one distinguishing
feature of this configuration is that it is 1/2-BPS for any
transverse position in $\alpha$ and its size is fixed.
However, we should notice from (\ref{d1diap}) that the dependence of
$\epsilon_{+\pm}$ on $\epsilon_{-\pm}$ changes continuously with
$\alpha$.

On the other hand, the configuration corresponding to the static
gauge $\zeta^2 = \varphi_-$ in (\ref{gaugediag}) leads to the
same result with that of (\ref{d1diap}) but with the replacement
$\Gamma_{35} \rightarrow - \Gamma_{35}$.  Therefore, it is also
1/2-BPS.

We now turn to the configurations where $\alpha$ is a worldvolume
coordinate not a transverse one.  Then another worldvolume coordinate is
along a circle embedded in the space composed of two S$^2$'s parametrized
by $\theta_{1,2}$ and $\phi_{1,2}$.  The possible
candidates for such circle are the circle given by
$\phi_1$ or $\phi_2$ and the `diagonal' one by $\varphi_+$ or
$\varphi_-$ defined in (\ref{diagvar}).  For the first case,
the static gauge is chosen to be
\begin{align}
\zeta^1 = \alpha \,, \quad
\zeta^2 = \phi_1 \,,
\label{hs1}
\end{align}
with constant $\theta_1$,
for which we have $\sqrt{|G|} = R^2 \cos \alpha \sin \theta_1$ and
\begin{align}
U =  e^{ -\frac{1}{2} \alpha \Gamma_{z1} }
  e^{ - \frac{1}{2} \theta_1 \Gamma_{z2} }
  e^{ \frac{1}{2} \phi_1 \Gamma_{23} }
\end{align}
from (\ref{usol}).  The next step is to compute
\begin{align}
\frac{1}{2 \sqrt{|G |} } U^{-1} \epsilon^{ij}
\gamma_{ij} \Gamma_z U
&= U^{-1} \Gamma_{13} \Gamma_z U
\notag \\
&=  \left(
        \cos \theta_1 e^{-\phi_1 \Gamma_{23}} + \sin \theta_1 \Gamma_{z2}
    \right) \Gamma_{13} \Gamma_z \,.
\end{align}
Although the last expression explicitly depends on the worldvolume coordinate
$\phi_1$, the dependence can be removed by taking $\theta_1 = \pi/2$,
corresponding to the great circle in S$^2$ parametrized by ($\theta_1$,
$\phi_1$).  At such position in $\theta_1$, Eq.~(\ref{susychk})
gives the supersymmetry structure of the configuration as
\begin{align}
\epsilon_{++}
&= \frac{\cos \beta}{1 + \sin \beta} z \Gamma_{123} \epsilon_{-+}
  - \Gamma_m x^m \epsilon_{--} \,,
\notag \\
\epsilon_{+-}
&= - \frac{\cos \beta}{1 - \sin \beta} z \Gamma_{123} \epsilon_{--}
  - \Gamma_m x^m \epsilon_{-+} \,,
\label{d1s1}
\end{align}
where $\epsilon_{+\pm}$ and $\epsilon_{-\pm}$ are defined in
(\ref{epm}) in terms of the projection operator (\ref{xproj}).
This result shows us that the configuration (\ref{hs1}) at
$\theta_1 = \pi/2$ is 1/2-BPS.  Similarly, we can confirm that
another configuration ($\zeta^1 = \alpha$, $\zeta^2 = \phi_2$)
at $\theta_2 = \pi/2$ is also 1/2-BPS.  Its supersymmetry structure
is also given by (\ref{d1s1}) but with the replacement
$\Gamma_{123} \rightarrow \Gamma_{z45}$.

Finally, as for the second configuration where $\varphi_+$ or $\varphi_-$
is a worldvolume coordinate, the corresponding static gauge is
($\zeta^1 = \alpha$, $\zeta^2 = \varphi_+$ or $\varphi_-$)
at constant $\theta_1$ and $\theta_2$.  However, this
configuration turns out to be non-supersymmetric since there is no way
to eliminate the dependence on the worldvolume coordinate
in the calculation of
$ U^{-1} \epsilon^{ij} \gamma_{ij} \Gamma_z U / 2 \sqrt{|G |}$
even by choosing certain values of $\theta_1$ and $\theta_2$.

\section{Evaluation of action and relation to $(p, q)$ string instanton}
\label{evalaction}

\begin{table}
\begin{center}
\begin{tabular}{c|cc}
\hline
   & ($\zeta^1$, $\zeta^2$) & position   \\
\hline\hline
\begin{tabular}{c} (i) \\ (ii) \\ (iii) \\ (iv) \end{tabular}
  & \begin{tabular}{c}
              ($\theta_1$, $\phi_1$)  \\
              ($\theta_2$, $\phi_2$)  \\
              ($\vartheta$, $\varphi_+$) \\
              ($\vartheta$, $\varphi_-$)
    \end{tabular} &
    \begin{tabular}{c}
               $\alpha = 0$ \\
               $\alpha = \pi/2$ \\
               any $\alpha$ \\
               any $\alpha$
    \end{tabular}
\\ \hline
\begin{tabular}{c} (v) \\  (vi) \end{tabular}
  &  \begin{tabular}{c}
                 ($\alpha$, $\phi_1$) \\
                 ($\alpha$, $\phi_2$)
     \end{tabular} &
     \begin{tabular}{c}
                 $\theta_1 = \pi/2 $ \\
                 $\theta_2 = \pi/2 $
     \end{tabular}
\\ \hline
\end{tabular}
\caption{\label{configs} 1/2-BPS D1-instanton configurations.
($\zeta^1$, $\zeta^2$): static gauge for a given
D1-brane configuration. position: transverse position. The definitions
for $\vartheta$ and $\varphi_\pm$ are given in (\ref{diagvar}).
All the configurations have the spherical shape with the radius
equal to the that of $S^5$, and are related by $SO(6)$ rotations,
the symmetry group of S$^5$. Thus any other configurations not listed
here are 1/2-BPS if they can be obtained by $SO(6)$ rotations
from, for example, the configuration (i).}
\end{center}
\end{table}

So far,  we have identified the 1/2-BPS instantonic D1-brane
configurations in the AdS$_5\times$S$^5$ background, which are
summarized in Table \ref{configs}.  Having the
configurations, if we evaluate the Euclidean action for all the
configurations listed in Table \ref{configs} by using the D1-brane action
(\ref{d1}) with $\sqrt{|G|}$'s computed for the configurations in the
last section, we see that the configurations from (i) through
(iv) have the same action value as\footnote{For the evaluation, the
explicit expressions for the AdS radius (\ref{adsrad}) and the
D1-brane tension (\ref{d1tension}) have been utilized.}
\begin{align}
S_{\mathrm{D1}}
= \frac{T}{\cos \beta} \int d^2 \zeta \sqrt{ | G | }
= \frac{T}{\cos \beta} \cdot 4 \pi R^2
= \frac{4}{\cos \beta} \sqrt{ \frac{\pi N}{g_s} } \,.
\label{s2value}
\end{align}
As for the remaining configurations, (v) and (vi), the action values
seem to be half of $S_{\mathrm{D1}}$ at first glance basically
because the range of $\alpha$ is $ 0 \le \alpha \le \pi/2$ and thus
the configurations look like hemispheres.  However,
this is not the case and the action values are still given by
$S_{\mathrm{D1}}$.  Let us explain the reason for the configuration
(v) by following the prescription of \cite{Hofman:2006xt} given in
a similar parametrization of S$^5$.  If we pick a pair of antipodal points
on S$^2$ parametrized by ($\theta_2, \phi_2$), which are
$\theta_2 = 0, \pi$ in the present case, then the pair together
with the coordinates $\alpha$ and $\phi_1$ form an
S$^2$.\footnote{We note that the coodinate $\theta_1$ is fixed at
$\pi/2$ for the configuration (v).}  This S$^2$
is the actual space that gives the shape of configuration (v).
Thus the configuration (v) has the action value of (\ref{s2value}),
not half of it, and importantly
its supersymmetry structure (\ref{d1s1}) is not spoiled for the
spherical shape.
The same process applies to the configuration (vi).
As a result, the whole configurations in Table \ref{configs} have
the spherical shape and have the action value of (\ref{s2value}).
Additionally, one notable fact is that they have the
maximum radius equal to that of S$^5$.  By using six Cartesian
coordinates, S$^5$ of unit radius is described as  $\sum_{i=1}^6 X_i^2=1$.
Associated with the S$^5$ metric (\ref{s5metric}), we may let
\begin{gather}
X_1 = \cos \alpha \cos \theta_1 \,, \quad
X_2 = \cos \alpha \sin \theta_1 \cos \phi_1 \,, \quad
X_3 = \cos \alpha \sin \theta_1 \sin \phi_1 \,,
\notag \\
X_4 = \sin \alpha \cos \theta_2 \,, \quad
X_5 = \sin \alpha \sin \theta_2 \cos \phi_2 \,, \quad
X_6 = \sin \alpha \sin \theta_2 \sin \phi_2 \,.
\label{cartesian}
\end{gather}
Then each configuration
appearing in Table \ref{configs} satisfies an algebraic equation
describing a sphere of unit radius:
$X_1^2 + X_2^2 + X_3^2 = 1$ for the configuration (i),
$X_4^2 + X_5^2 + X_6^2 = 1$ for (ii),
 $\sum_{i=1}^6 X_i^2=1$ for (iii) and (iv),\footnote{The
algebraic equation actually represents a sphere not S$^5$ because
$\alpha$ is fixed at a certain value and angles are identified as
(\ref{diagvar}).}
$X_2^2 + X_3^2 + X_4^2 = 1$ for (v), and
$X_1^2 + X_5^2 + X_6^2 = 1$ for (vi).  This means that all
the configurations  appearing in Table \ref{configs} are related by
$SO(6)$ rotations, the symmetry group of S$^5$.

After obtaining the results, one can see that it is enough to find just one 1/2-BPS
configuration, for example the configuration (i), because other 1/2-BPS configurations are obtained by $SO(6)$ rotations. However, the
investigation of supersymmetry structure would require additional
task related to the analysis of associated rotated spinors.
Starting from Table \ref{tablebps}, we got the information that
1/2-BPS D1-instanton should wrap two cycle in S$^5$.
But we do not know the geometry of the two cycle, a priori.
The geometry is automatically determined when we analyze the
1/2-BPS condition.  Before we produce Table \ref{configs},
it is not clear whether the other geometry than the sphere is
excluded.  But we came to know that the only 1/2-BPS configuration
is the sphere with the maximal radius.
On the other hand, our analysis leads to the torus like configuration
(\ref{t2}), but it turns out not to be supersymmetric.
If it were 1/2-BPS, we would
have different class of configurations with toroidal shape.
Even a spherical shape configuration is 1/2-BPS only if it has
the certain radius, the radius of S$^5$.
Thus, based on our results, it is concluded that any
spherical D1-instanton with the radius
equal to that of S$^5$ is 1/2-BPS.\footnote{At present, it is not clear
if there are other 1/2-BPS configurations with nontrivial
topology such as surface with genus greater than one.}

Since the 1/2-BPS D1-instantons have the same shape and size and are
related to each other by $SO(6)$ transformations, one may
think that the configurations listed in Table \ref{configs} are
redundant and only one of them is important.  This is definitely
the case if we are only interested in single D1-instanton.
However, if we want to study the
supersymmetric multiple D1-instantons, the configurations in
Table \ref{configs} with their supersymmetry structures are necessary
and important.  We will discuss this issue in the next section.

The action value of (\ref{s2value}) allows us to estimate the saddle-point
contribution of D1-instanton to any physical process.  For such
estimation, it is instructive to rewrite the D1-instanton action
in terms of the well known D(-1)-brane or D-instanton action in the
AdS$_5\times$S$^5$ background
\cite{Chu:1998in, Kogan:1998re, Bianchi:1998nk}. Namely, by using the
D(-1)-brane action given by
\begin{align}
S_{\mathrm{D(-1)}} = \frac{2 \pi}{g_s} = \frac{8 \pi^2 N}{\lambda} \,,
\end{align}
where $\lambda = 4 \pi g_s N$ is the usual 'tHooft parameter,
let us rewrite (\ref{s2value}) as
\begin{align}
S_{\mathrm{D1}}
= \frac{\sqrt{\lambda}}{\pi} S_{\mathrm{D(-1)}} \,,
\label{d1vsd-1}
\end{align}
where the worldvolume gauge field strength has been turned off for
simplicity ($\beta = 0$).  Because the D(-1)-brane action is
proportional to $N$ and thus its contribution is suppressed by $e^{-N}$,
it is now clear that the contribution of D1-instanton also leads to the
same suppression factor.  If we turn on the worldvolume gauge
field strength and increase it, the suppression becomes much stronger
as one can see from (\ref{s2value}).  In this sense, the D1-instanton
with vanishing or weak worldvolume gauge field is preferable.

Until now, we have evaluated the D1-instanton action.  By the way,
the D1-brane action in (\ref{s2value}) (or (\ref{d1})) has the form
of fundamental string action.  Since it is well known that the D1-brane
is related to the fundamental string or more generally $(p, q)$ string
through the $SL(2, \mathbf{Z})$ transformation \cite{Schwarz:1995dk,Witten:1995im}, one may be curious as to how
the action of (\ref{d1}) is written in an $SL(2, \mathbf{Z})$
covariant way.  For this, we consider a D1-brane and turn
on constant non-zero RR 0-form or axion field $C_{(0)}$ which is allowed
in the AdS$_5\times$S$^5$ background.  From (\ref{action}), the
corresponding action is given by
\begin{align}
S = \int d^2 \zeta \mathcal{L}
  =  T \int d^2 \zeta \sqrt{ | G + \mathcal{F} | }
   - i  g_s T \int \chi \mathcal{F} \,,
\label{gend1}
\end{align}
where the constant axion has been denoted as $\chi$.

Although the relation between the action (\ref{gend1}) and that of
$(p, q)$ string instanton could be obtained simply by Euclideanizing
the result of \cite{Oda:1998ad,Park:1998uj}, we will show it directly in
Euclidean space by following the process described in \cite{Drukker:2011zy}
and let $q=1$ since single D1-brane action is considered.
The flux density $p$ of worldvolume gauge field strength interpreted
as the number of fundamental strings dissolved into D1-brane is
the momentum dual to the gauge field strength $F_{12}$,
\begin{align}
p &= - i \frac{\partial \mathcal{L}}{\partial F_{12}}
\notag \\
  &= -i \frac{\mathcal{F}_{12}}{g_s  \sqrt{ | G + \mathcal{F} | } }
     - \chi \,.
\label{flux}
\end{align}
This flux density is surely a constant because of the equation of
motion (\ref{eom})\footnote{The presence of constant axion $\chi$
does not modify the equation of motion (\ref{eom}) because the
second term on the right hand side of (\ref{gend1}) becomes
a topological one for constant $\chi$.} and the constant $\chi$.
On the other hand,
$F_{12}$ depends on the worldvolume coordinates as can be seen from
Eqs.~(\ref{bdef}) and (\ref{emsol}).  Thus it would be appropriate to
express the action in terms of $p$ rather than $F_{12}$.  If we denote
such action as $S_{(p,1)}$, it is related to $S$ of (\ref{gend1})
through the Legendre transformation as follows:
\begin{align}
S_{(p,1)}
&= S -  \int d^2 \zeta i p F_{12}  \notag \\
&= \frac{1}{2\pi \alpha'} \sqrt{ (p + \chi)^2 + \frac{1}{g_s^2} }
   \int d^2 \zeta \sqrt{ | G | } \,.
\end{align}
We see that this result is nothing but the Nambu-Goto
string instanton action with the $SL(2, \mathbf{Z})$ covariant
tension
\begin{align}
T_{p,1}
= \frac{1}{2\pi \alpha'} \sqrt{ (p + \chi)^2 + \frac{1}{g_s^2} } \,,
\end{align}
that is, the $(p, 1)$ string instanton action.
From this relation, we see that the D1-instanton action
(\ref{s2value}) is related to $S_{(p,1)}$ with
\begin{align}
\cos \beta = \frac{1}{\sqrt{g_s^2 (p + \chi)^2 + 1}} \,,
\end{align}
where the constant axion $\chi$ has been turned on.

\section{Discussion}
\label{concl}

The single 1/2-BPS D1-instanton has been of our interest. As a next
step, we may ask if multiple D1-instanton configuration is
supersymmetric or not.
The simplest case would be that of two D1-instantons of the
same type, one of six types of configurations in Table \ref{configs},
at different positions in the AdS$_5$ space. However, it turns out
that such D1-instanton configuration breaks all the supersymmetry.
For example, let us take two D1-instantons of the configuration (i)
in Table \ref{configs} as a representative whose positions in the AdS$_5$
space are $(z_{(1)}, x^m_{(1)})$ and $(z_{(2)}, x^m_{(2)})$ respectively.
Then we have two sets of equations from (\ref{d1s2res});
\begin{align}
\epsilon_{++}
&=  z_{(1)} \Gamma_{23} \Gamma_z \epsilon_{-+}
  - \Gamma_m x_{(1)}^m \epsilon_{--} \,,
\notag \\
\epsilon_{++}
&=  z_{(2)} \Gamma_{23} \Gamma_z \epsilon_{-+}
  - \Gamma_m x_{(2)}^m \epsilon_{--} \,,
\notag \\
\epsilon_{+-}
&= -  z_{(1)} \Gamma_{23} \Gamma_z \epsilon_{--}
  - \Gamma_m x_{(1)}^m \epsilon_{-+} \,,
\notag \\
\epsilon_{+-}
&= -  z_{(2)} \Gamma_{23} \Gamma_z \epsilon_{--}
  - \Gamma_m x_{(2)}^m \epsilon_{-+} \,,
\end{align}
where the worldvolume gauge field strength has been turned off
($\beta = 0$) because it does not play any crucial role for the
consideration of supersymmetry.
We note that the two $\epsilon_{++}$'s ($\epsilon_{+-}$'s) on the left
hand side should be the same if some fraction of supersymmetry
is preserved.  This allows us to get two equations by subtracting
the first (last) two equations;
\begin{align}
0 &= \left(z_{(1)} - z_{(2)} \right) \Gamma_{23} \Gamma_z \epsilon_{-+}
  - \Gamma_m \left(x_{(1)}^m - x_{(2)}^m \right)\epsilon_{--} \,,
\notag \\
0 &=  - \left(z_{(1)} - z_{(2)} \right) \Gamma_{23} \Gamma_z \epsilon_{--}
  - \Gamma_m \left(x_{(1)}^m - x_{(2)}^m \right) \epsilon_{-+} \,.
\end{align}
By using these, one can check supersymmetry for the three cases,
($z_{(1)} \neq z_{(2)}$, $x_{(1)}^m = x_{(2)}^m$),
($z_{(1)} = z_{(2)}$, $x_{(1)}^m \neq x_{(2)}^m$), and
($z_{(1)} \neq z_{(2)}$, $x_{(1)}^m \neq x_{(2)}^m$).  However,
as one can see easily, none of them are supersymmetric.
Thus, two or more D1-instantons of the same type are supersymmetric
(1/2-BPS) only if they are coincident in the AdS$_5$ space.
This is in contrast to the D(-1)-brane case where multiple D(-1)-branes
do not spoil the supersymmetry structure even if they are apart.

Some more generalized case is that of two different types of
D1-instantons. From the Table \ref{configs}, we can think of fifteen
combinations.  However, we will not investigate all of them but instead
intend to show just the possibility for the existence of supersymmetric
configuration by taking one example.  It is the combination composed
of D1-instantons (i) and (ii).  If the D1-instantons are taken to be
coincident in the AdS$_5$ space\footnote{Actually, like in the previous
case, they do not form a supersymmetric object if they
are separated in the AdS$_5$ space.} and the worldvolume gauge field
strengths on both of them are turned off for simplicity,
Eq.~(\ref{d1s2res}) for (i) becomes
\begin{align}
\epsilon_{++} = z \Gamma_{23} \Gamma_z \epsilon_{-+} \,, \quad
\epsilon_{+-} = -  z \Gamma_{23} \Gamma_z \epsilon_{--} \,,
\label{d1-1}
\end{align}
and the corresponding equation for (ii) is
\begin{align}
\epsilon_{++} = z \Gamma_{45} \Gamma_z \epsilon_{-+} \,, \quad
\epsilon_{+-} = -  z \Gamma_{45} \Gamma_z \epsilon_{--} \,,
\label{d1-2}
\end{align}
which follows from (\ref{d1s2res}) with the replacement
$\Gamma_{23} \rightarrow \Gamma_{45}$ as mentioned below
(\ref{d1s2res}).  Because $\Gamma_{23}$ and $\Gamma_{45}$
commute with each other,\footnote{They also commute with $\hat{\gamma}$
and $\Gamma^{11}$ measuring the ten dimensional chirality.}
it is convenient to split $\epsilon_{ab}$ ($a, b = \pm$) in terms of
their eigenvalues as $\epsilon_{abss'}$;
\begin{align}
i\Gamma_{23} \epsilon_{ab \pm s'}  = \pm \epsilon_{ab \pm s'} \,, \quad
i\Gamma_{45} \epsilon_{ab s \pm}  = \pm \epsilon_{ab s \pm} \,.
\label{extproj}
\end{align}
Although, generically, the constant spinors $\epsilon_{++}$
($\epsilon_{+-}$) appearing in (\ref{d1-1}) and (\ref{d1-2}) have
different dependences on $\epsilon_{-+}$ ($\epsilon_{--}$), they should
be the same if the present D1-instanton combination is supersymmtric.
If we now subtract the first (second) equation of
(\ref{d1-2}) from that of (\ref{d1-1}) and project the resulting equation
according to the eigenvalues of $\Gamma_{23}$ and $\Gamma_{45}$
by using (\ref{extproj}), then we get four equations as
\begin{align}
0 = 2 i z \Gamma_z \epsilon_{-++-} \,, \quad
0 = - 2 i z \Gamma_z \epsilon_{-+-+} \,,
\notag \\
0 = - 2 i z \Gamma_z \epsilon_{--+-} \,, \quad
0 = 2 i z \Gamma_z \epsilon_{---+} \,.
\end{align}
This shows that the four constant spinors on the right hand
sides, one half of $\epsilon_{-\pm}$, should vanish for generic $z$
position. In the end, the following ones, another half of
$\epsilon_{-\pm}$, remain free parameters,
\begin{align}
\epsilon_{-+++} \,, \quad \epsilon_{-+--} \,, \quad
\epsilon_{--++} \,, \quad \epsilon_{----} \,,
\end{align}
and thus the combination of D1-instantons (i) and (ii) coincident
in the AdS$_5$ space turns out to be 1/4-BPS.

Having an explicit example of supersymmetric multiple D1-instatons,
we may expect other possibilities from the remaining two D1-instanton
configurations. However we will not pursue them further since our
primary concern is single 1/2-BPS D1-instanton. We hope to have an opportunity to deal with them in a near future.


\section*{Acknowledgments}

This research was supported by Basic Science Research Program through the
National Research Foundation of Korea (NRF) funded by the Ministry of
Education with Grant No.~NRF-2018R1A2B6007159, NRF-2021R1A6A1A10042944 and NRF-2021R1A2C1012440 (JP),
and NRF-2018R1D1A1B07045425 (HS).





\providecommand{\href}[2]{#2}\begingroup\raggedright\endgroup

\end{document}